\documentclass[aps,prb,twocolumn,showpacs,amsmath,amssymb,superscriptaddress]{revtex4-1}

\usepackage{graphicx}% Include figure files
\usepackage{dcolumn}% Align table columns on decimal point
\usepackage[dvipsnames]{xcolor}

\usepackage{bm}% bold math
\usepackage{color}
\usepackage{graphicx}
\usepackage{epsfig}
\usepackage{fancyhdr}
\include{graphic}

\setcitestyle{numbers,square}

\begin{document}

\title{Microscopic mechanism of level attraction}

\author{Bimu~Yao}
\affiliation{State Key Laboratory of Infrared Physics, Shanghai Institute of Technical Physics, Chinese Academy of Sciences, Shanghai 200083, People's Republic of China}
%\affiliation{Department of Physics and Astronomy, University of Manitoba, Winnipeg, Canada R3T 2N2}

\author{Tao~Yu}\email{T.Yu@tudelft.nl}
\affiliation{Kavli Institute of NanoScience, Delft University of Technology, 2628 CJ Delft,  The Netherlands}

\author{Xiang~Zhang}
\affiliation{Kavli Institute of NanoScience, Delft University of Technology, 2628 CJ Delft,  The Netherlands}

\author{Wei~Lu}
\affiliation{State Key Laboratory of Infrared Physics, Shanghai Institute of Technical Physics, Chinese Academy of Sciences, Shanghai 200083, People's Republic of China}

\author{Yongsheng~Gui}
\affiliation{Department of Physics and Astronomy, University of Manitoba, Winnipeg, Canada R3T 2N2}

\author{Can-Ming~Hu}
\affiliation{Department of Physics and Astronomy, University of Manitoba, Winnipeg, Canada R3T 2N2}

\author{Yaroslav~M.~Blanter}
\affiliation{Kavli Institute of NanoScience, Delft University of Technology, 2628 CJ Delft,  The Netherlands}

\date{\today}

\begin{abstract}

The emerging level attraction from dissipative light-matter coupling converges the typical Rabi-splitting feature from coherent coupling and exhibits potentials in topological information processing. However, the underlying microscopic quantum mechanism of dissipative coupling still remains unclear, which brings difficulties in quantifying and manipulating coherence-dissipation competition and thereby the flexible control of level attraction. Here, by coupling magnon to a cavity supporting both standing and travelling waves, we identify the travelling-wave state to be responsible for magnon-photon dissipative coupling. By characterizing radiative broadening of magnon linewidth, we quantify the coherent and dissipative coupling strengths and their competition. The effective magnon-photon coupling strength, as a net result of competition, is analytically presented in quantum theory to show good agreement with measurements. In this manner, we extend the control dimension of level attraction by tuning field torque on magnetization or global cavity geometry. Our finding opens new routines to engineer coupled harmonic oscillator system.

\end{abstract}

%\keywords{Spintronics, Polariton, Ferromagnetic Resonance}

%\pacs{85.75.-d, 71.36.+c, 76.50.+g}
\maketitle

\section{INTRODUCTION}

The control of resonant interactions between light and matter offers opportunities to exchange information among different entities, and thereby has received significant attention \cite{Haroche,Re,PTGS,scalable}. Conventionally, an elementary excitation of matter can strongly and coherently couple to a well-isolated electromagnetic environment to form two new hybridized states showing repulsion of energy levels, i.e., Rabi splitting \cite{semirabi}, which is at the heart of quantum information processing \cite{semirabi,Networks,Transport} and coherent manipulation between distinct quantum objects \cite{Huebl2013, Tabuchi2015, zhang2016, Bai2015}. In realistic situations, quantum coherence is complemented by dissipation, which needs to be accounted for in order to understand the behaviour of coupled quantum systems. Dissipation can give rise to non-Hermitian quantum dynamics leading to new physical phenomena such as super- and sub-radiance \cite{subradiance1,subradiance2,subradiance6}, non-Hermitian skin effect \cite{non_hermitian0,non_hermitian9,non_hermitian5}, topological non-Hermitian phases \cite{non_hermitian5,non_hermitian6}, and critical behaviour beyond the standard paradigms \cite{D1,non_hermitian3,D3}. Particularly, through dissipation engineering, the nature of the coupling between energy levels can be dramatically modified, even demonstrating level attraction when the frequencies of the two coupled modes match \cite{MH18,XK,BR,Stamps}. Such attraction opens perspectives for topological energy transfer, quantum sensing, and synchronization in hybrid systems \cite{XK,MH18,YY,BR}. 

To understand the origin of the level attraction and to facilitate its control, it is fundamentally important to distinguish between the coherent and the dissipative coupling strengths through measurement, and particularly to directly measure the dynamics caused by coherent and dissipative coupling. We note that these issues can not be simply solved through measuring the level splitting gap, which is a combined effect from the coherent and dissipative couplings. Recently, experimental realizations of the level attraction in one-dimensional (1D) Fabry-Perot-like cavity \cite{MH18} and inverted pattern of split-ring resonator \cite{SSR,YY} between the yttrium iron garnet (YIG) magnon and the cavity photon provided a path to explore this question. To explain the experimentally realized level attraction, classical theory with phenomenological parameters were proposed, including the cavity Lenz effect, analogue of circuits, and the dual driving forces on the magnon\cite{MH18,SSR,YY,XK}. These interpretations cannot provide microscopic origin of the magnon-photon dissipative coupling, nor do they give quantitative understanding on the competition between coherence and dissipation. We note that the systems used in the experiments \cite{MH18,SSR,YY} are often open and coupled to the environment, allowing the magnon to radiate out energy by travelling photon, which is hence dissipative in nature \cite{DOS}. The existence of travelling modes and their effect on magnon-photon hybridization have not been sufficiently explored to understand the phenomenon of the level attraction.

In this work, we demonstrate that the travelling photon wave is the microscopic origin that causes the dissipative magnon-photon coupling, which is responsible for the generation of level attraction dynamics. Coupling with the magnon mode in a YIG sphere, the travelling waves cause the dissipation of magnon through radiating the energy to open environment, which is different from the intrinsic dissipations of magnon and photon themselves, and promise new perspective in engineering the dissipation-coherence competition. We present an experimental method and a theoretical model to distinguish between the coherent and dissipative couplings in a prototype open quantum system, e.g. 1D Fabry-Perot-like cavity, which supports both standing and travelling (continuous) wave states. By measuring the radiative broadening of magnon linewidth at and away from the cavity resonance, we are able to determine the coupling strengths in the coherent and dissipative nature, respectively, and reveal that level attraction (repulsion) arises when the dissipative (coherent) coupling strength is dominant. This is yet entirely unforeseen in coupled magnon-photon dynamics. The resultant effective coupling strength obtained from photon-transmission measurement is found to be an outcome of dissipation-coherence competition, instead of coherence alone. The interaction between magnon and standing or travelling photon modes can be adjusted by tuning the torque exerted on the magnetization by the microwave magnetic field, or by tuning the global cavity geometry, allowing us to manipulate the competition between the coherence and dissipation at will, and hence flexibly control level attraction. In this sense, our experiment is, to our knowledge, unique for providing a simple method to quantitatively highlight the differences between coherence and dissipation, as well as distinctive from existing works by extending the control dimensions of level attraction. We note that, since our mechanism is based on the harmonic oscillators, it can be applied to other harmonic systems including on-chip integrated circuits or optomechanical devices, and can be readily extended to the quantum regime at millikelvin temperatures with single magnon excitation \cite{onemagnon,dark,RA}.

%we present an experimental method and a theoretical model to distinguish between the coherent and dissipative \textcolor{blue}{magnon-photon} couplings, and explain the level attraction by magnon radiation in a prototype open quantum system, e.g. 1D Fabry-Perot-like cavity, in which both standing and travelling (continuous) wave states exist. By measuring the magnon radiative linewidth at and away from the cavity resonance due to the magnon radiation, we can determine the relative coupling strengths in the coherent and dissipative nature, respectively. We reveal that, level attraction (repulsion) arises when the dissipative (coherent) coupling strength is dominant in the competition. The resultant effective coupling strength obtained from transmission spectra of photon is found to be an outcome of dissipation-coherence competition, instead of coherence alone. The interaction between magnon and standing or travelling photon modes can be adjusted by tuning the torque exerted on the magnetization by the microwave magnetic field, or by tuning the global cavity geometry, allowing us to manipulate the competition between the coherence and dissipation, and hence flexibly control level attraction.  We note that, since our mechanism is based on the harmonic oscillators, it can be applied to other harmonic systems including on-chip integrated circuits or optomechanical devices, and can be readily extended to the quantum regime at millikelvin temperatures with a single magnon excitation \cite{onemagnon,dark,RA}. 

\section{Field-torque induced level attraction}

We used a 1D Fabry-Perot-like cavity with two rectangular-to-circular transitions at the two terminals and a 16~mm-diameter circular waveguide in the middle \cite{hmode} to couple to the magnon Kittel mode in a YIG sphere, as sketched in Fig.~\ref{Fig1}(a). The co-polarized $\rm TE_{11}$ mode in the circular waveguide can be divided into two components \cite{yaoapl}. One component has polarization consistent with the fundamental mode of rectangular port, enabling the travelling waves to be transmitted through with almost zero reflection. While the rest component is totally reflected from both terminals, forming the standing waves. Thus, similar to the acoustics physics in a \textit{``flute}", our waveguide cavity readily support the standing waves at $\omega_c$ and travelling waves at detuned frequencies. Here we performed the measurements around the mode with an antinode in the microwave magnetic field at the center of cavity, with the cavity mode frequency $\omega_c/2\pi=12.14~\rm GHz$ and a damping coefficient of $9\times10^{-3}$. By introducing an 1-mm in diameter highly-polished and low-damping ($\sim 10^{-5}$) YIG sphere to the cavity middle plane, our device enables the magnon mode to interact with both the standing and travelling microwaves in a broadband range.

\begin{figure} [!htbp]
	\begin{center}

		\epsfig{file=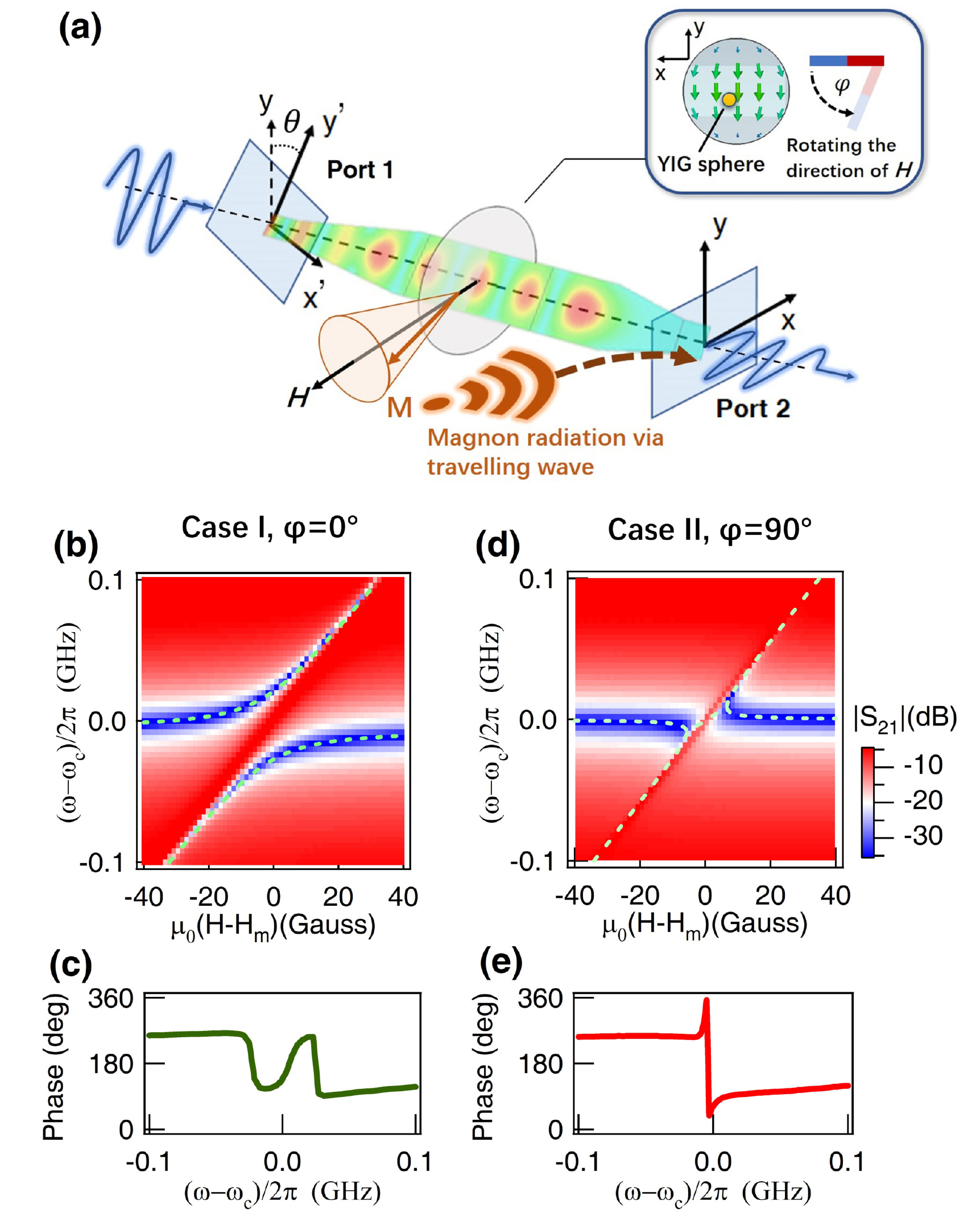,width=8.8cm}
		\caption{Experimental observation of level repulsion and attraction. (a) Sketch of experimental configuration, the waveguide cavity supports standing wave at resonant cavity mode and travelling wave at detuned frequency, respectively. A YIG sphere is coupled to our cavity with its magnetization ($M$) biased to the direction of external magnetic field ($H$). Distribution of microwave magnetic field ${\bf h}$ is shown in the inset, with its relative orientation with respect to $H$ can be tuned by the angle $\varphi$. For case I ($\varphi=0^{\circ}$) and case II ($\varphi=90^{\circ}$), by measuring the photon transmission spectra for different $H$, we display amplitude map of level repulsion (attraction) in (b) [(d)], with the phase spectra at zero detuning shown in (c)[(e)]. } 
		\label{Fig1}
	\end{center}
\end{figure}

For generating various hybrid dynamics, the magnon-photon coupling is tuned by changing the direction of external magnetic field $H$. We consider here two typical situations. In case I, $H$ is set to be perpendicular to the polarization of the microwave magnetic field ${\bf h}$ at $\omega_c$ to maximize the cavity-field torque on the magnetization \cite{TQ}, and thereby the largest coherent coupling strength $\tilde{g}_s$ that is defined by the coupling between magnon and standing-wave component of photon \cite{IEEE} [see schematic figure in the inset of Fig.~\ref{Fig1}(a) with the distribution of microwave magnetic field at $\omega_c$, obtained from numerical simulation]. To pick up the response of the coupled states, the two ports of the waveguide cavity are connected to a vector network analyzer (VNA) for measuring the transmission spectra. By tuning to the magnon resonant field $H_m$, magnon mode frequency $\omega_{\rm K}$ is accordingly tuned to match $\omega_c$. Level repulsion feature in Fig.~\ref{Fig1}(b) is observed, indicating a coherent energy transfer between magnons and photons. Besides, the phase of the transmission at zero detuning is plotted in Fig.~\ref{Fig1}(c), in which the phase spectra displays typical behaviour of the level repulsion with two separated $\pi$-phase jumps, corresponding to the two ``splitting" modes in the level repulsion \cite{Lineshape}. 

We subsequently tuned the orientation of $H$ to suppress the coherent dynamics. Typically, by rotating $H$ by an angle of $\varphi$=$90^{\circ}$ (case II), the cavity-field torque on the magnetization is minimized to be zero, which seemingly is not able to drive magnon dynamics. This would indicate that the photon mode is not coupled to the magnon mode, and photon transmission would only show a bare cavity profile, which is independent of $H$. Whereas such a statement is true for coupling magnon to a well-confined cavity \cite{IEEE}, we show here that in our waveguide cavity this minimization of the field torque leads to the novel type of dispersion---level attraction. The photon transmission, shown in Fig.~\ref{Fig1}(d), displays the characteristic energy spectra of level attraction, with the resonance frequencies of the two modes attracted and converged to a single level around the resonance $\omega_{\rm K}$=$\omega_c$. The observation of level attraction has been further verified by the phase jump of the transmission at zero detuning in Fig.~\ref{Fig1}(e). In contrast to the level repulsion, we observed an approximate $2\pi$ phase jump at zero detuning instead, which results from the coalescence of the magnon mode and the photon mode \cite{MH18,YY}.

\section{Measuring the coherent and dissipative coupling strength}
The above observation indicates that besides the common standing wave in a cavity that causes coherent magnon-photon coupling, there are other ingredients causing the dissipative magnon-photon coupling that is crucial to the generation of level attraction \cite{MH18}. We note that our waveguide cavity also supports travelling waves when detuned from resonant frequencies. This can be directly confirmed by measuring transmission $|S_{21}|$ without applying $H$, as shown in Fig.~\ref{Fig2}(a). Trapped standing waves induce a sharp dip in the transmission spectra at $\omega_c$, while travelling waves deliver energy from one port to the other,  and cause an approximate unity transmission at detuned frequencies. Around $\omega_c$, the photonic mode is actually a complex superposition of standing and travelling waves. In this manner, our above intuition developed on the basis of the coupling between a magnon and a pure standing wave photon fails, which is not able to describe the results of our experiment.

\begin{figure} [!htbp]
	\begin{center}
		\epsfig{file=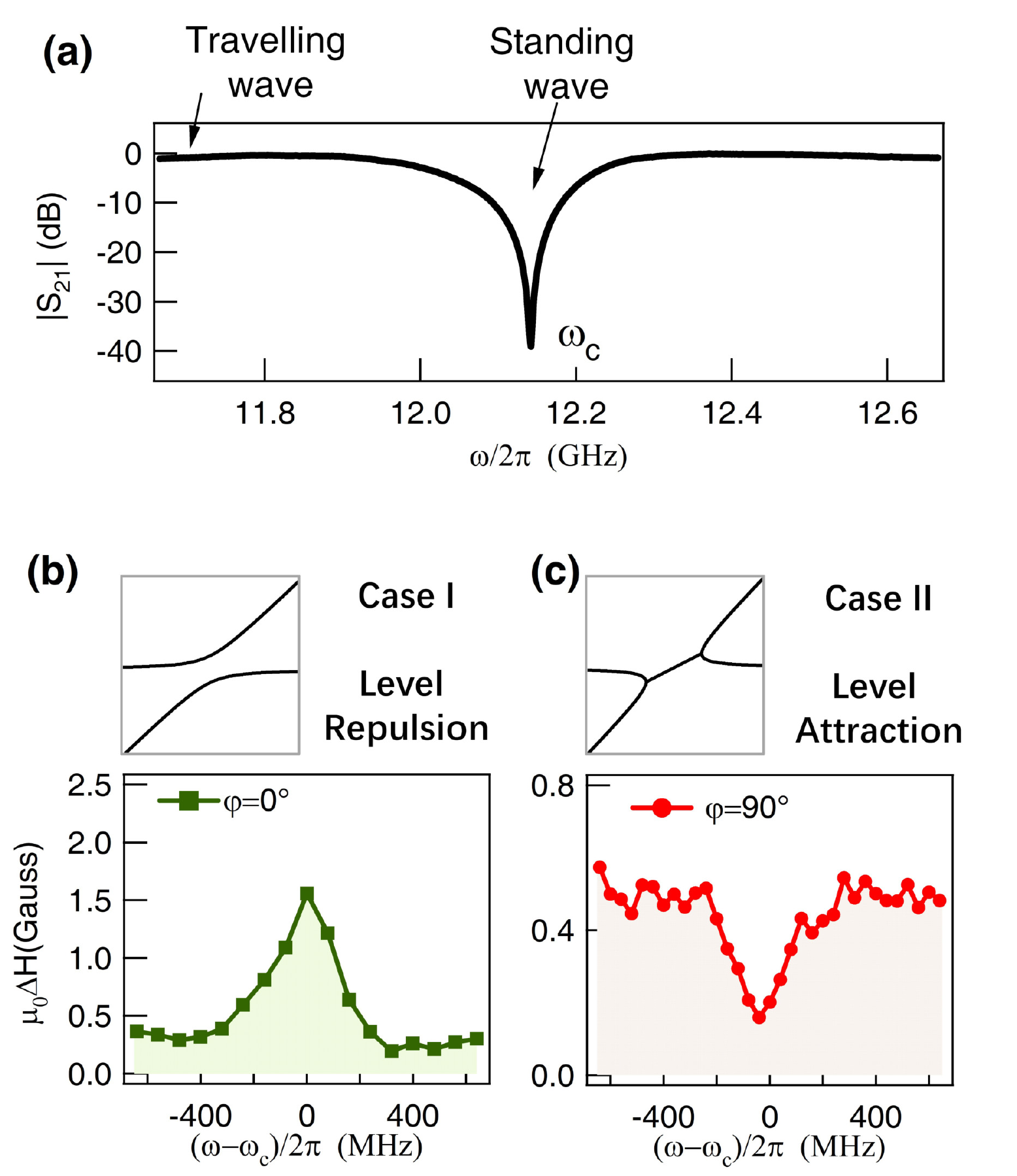,width=8.5cm}
		\caption{Magnon radiative linewidth. (a) Microwave transmission $|S_{21}|$ for our circular waveguide cavity with zero static magnetic field applied. At cavity resonance ($\omega_c$), our cavity supports standing waves, while at detuned frequencies, our cavity supports the travelling waves to deliver energy to the environment. Magnon radiative linewidth spectra are displayed for level repulsion case with $\varphi=0^{\circ}$ (b) and level attraction case with $\varphi=90^{\circ}$ (c), respectively.} 
		\label{Fig2}
	\end{center}
\end{figure}

Therefore, we need to go one step further to investigate the role of travelling waves. Continuous background, which is the travelling-wave component of photon, provides the dissipative nature in the coupling because a magnon can directly radiate out a photon that leaves the cavity and does not exert any backaction on the magnon. Rates of magnon radiation are determined by the square of magnon-photon coupling strength (details shown in theory part below). Therefore, by characterizing the magnon radiation rates via the measurement of the radiative linewidth $\mu_0\Delta H$, we can obtain the information of the coupling strengths between the magnons and standing (travelling) wave photons. Specifically, when detuned away from $\omega_c$, the coupling $\tilde{g}_b$ between the magnon and continuous background of photon determines the radiative linewidth with $\mu_0\Delta H=\pi\tilde{g}_b^2/\gamma$ with $\gamma $ being the gyromagnetic ratio. While at $\omega_c$, the standing waves bring coherent coupling $\tilde{g}_s$ to superpose with $\tilde{g}_b$, leading the radiative linewidth to be $\pi(\tilde{g}_s+\tilde{g}_b)^2/\gamma$ instead of $\pi\tilde{g}^2_s/\gamma$. The rigorous formulations and explanations are referred to the theoretical model part below. This understanding provides an effective method to compare the coherent and dissipative magnon-photon couplings ($\tilde{g}^2_{s}$ and $\tilde{g}^2_{b}$).
It is understood that when the radiative linewidth becomes the same at and detuned from the resonance, which is defined as the exchange point, the coherent coupling $\tilde{g}_s$ becomes zero and the magnon radiation is determined by the travelling waves.

For case I with level repulsion, the magnon radiative linewidth $\mu_0\Delta H$, which is obtained by subtracting the contribution from the intrinsic Gilbert damping and inhomogeneous broadening \cite{H0} in the measured total linewidth, displays an enhancement around $\omega_c$ in Fig.~\ref{Fig2}(b), revealing that the coherence is much greater than dissipation\cite{supp} (see detailed measurements in Supplementary Sections~1 and 2). Such a dominant coherent coupling effect is a typical behaviour shared by most coherently coupled systems, and thereby we obtained a canonical feature with level repulsion. Strikingly, different from such linewidth enhancement, for case II we found the enhancement is reversed to a suppression around $\omega_c$, suggesting $(\tilde{g}_s+\tilde{g}_b)^2<\tilde{g}_b^2$. This relation provides the information that $\tilde{g}_s$ has the opposite sign compared to the dissipative coupling $\tilde{g}_b$ as well as the magnitude of the dissipative coupling overcomes the coherent one, i.e., $|\tilde{g}_b|>|\tilde{g}_s|$. Such relative linewidth broadening at detuned frequencies clarifies the role of the travelling wave states: They radiate the magnon energy to the environment, causing dominant dissipative coupling which is essential to level attraction. Further, by engineering the travelling wave states, we demonstrate in the next section that relative strength between coherent and dissipative coupling can be gradually tuned, and a transition between level repulsion and attraction can be realized.

\section{Transition from repulsion to attraction via dissipation engineering}
For a better understanding of dissipative nature from the travelling waves, we combine the ``field-torque control technique"  described above and the ``magnon radiative linewidth characterization" (I) to continuously control the evolution of both coherent and dissipative coupling, (II) to observe the transition between repulsion and attraction.

(I) To investigate the competition between dissipative and coherent dynamics, we gradually turn $\varphi$ from $0^{\circ}$ to $90^{\circ}$. Varying the orientation of the static magnetic field $H$ adjusts the component of the microwave magnetic field that is orthogonal to $H$. In the measurement, we found that, as a function of $\varphi$, the transition between the standing-wave dominant regime and the travelling-wave dominant regime can be clearly observed in magnon radiative linewidth in Fig.~\ref{Fig3}(d). These observations indicate that competition between coherence and dissipation can be classified as follows.
 
 \begin{figure} [t]
 	\begin{center}
 		\epsfig{file=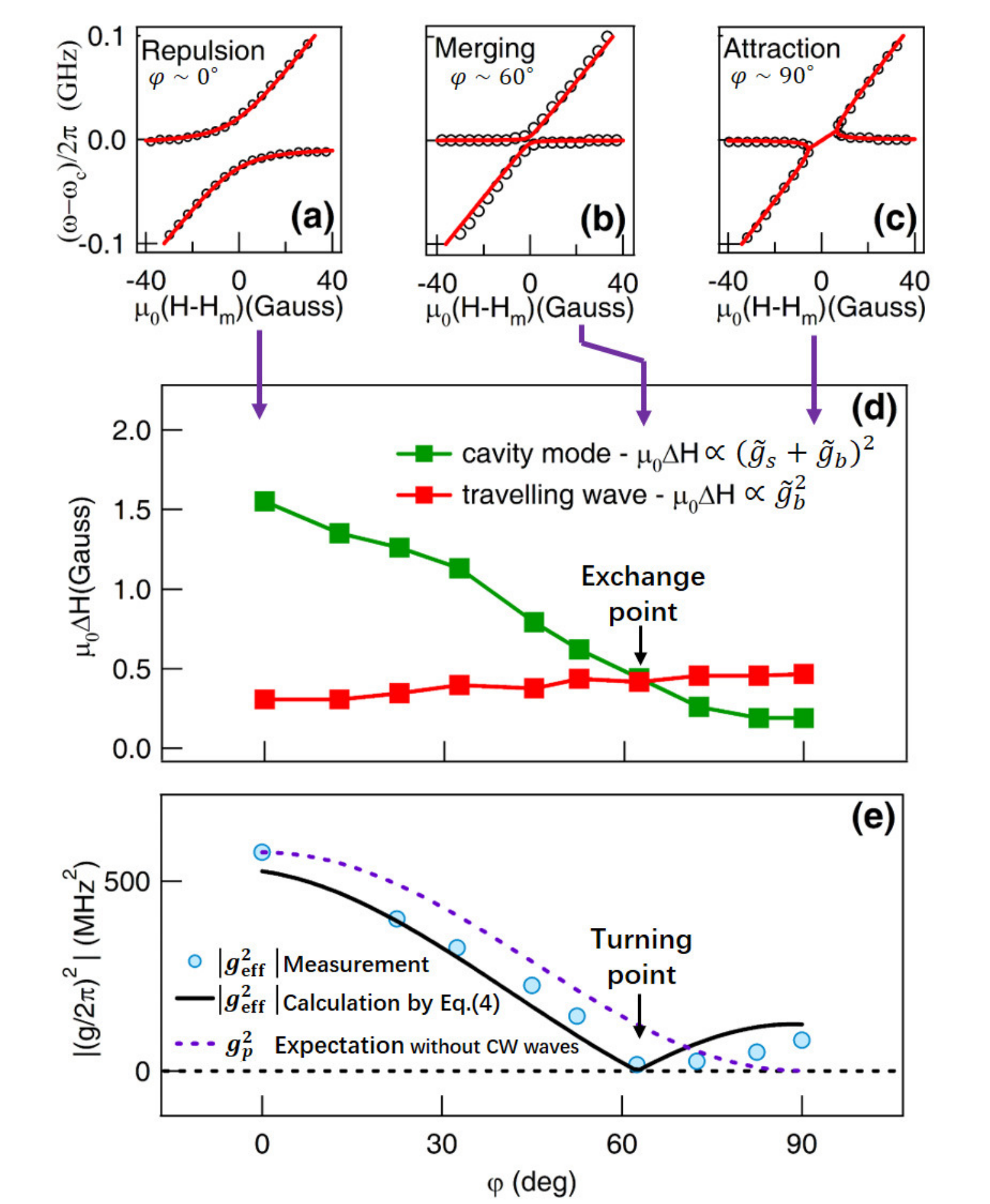,width=8.8cm}
 		\caption{Transition to level attraction via coherence-dissipation competition. Dispersion of coupled magnon-photon states measured by photon transmission, with level repulsion (a), level merging (b) and level attraction (c). When tuning $\varphi$, magnon radiative linewidth due to the standing (at cavity resonance) and travelling (about 400~MHz away from cavity resonance) wave is displayed in (d), with turning point suggesting the transition between repulsion and attraction. (e) Measured and calculated effective coupling strengths as function of $\varphi$ are plotted in blue circles and black solid lines, respectively. Purple dashed line shows the expected coupling strength without considering travelling waves.}
 		\label{Fig3}
 	\end{center}
 \end{figure}

In Fig.~\ref{Fig3}(d), in the first regime with $\varphi$ turned from $0^{\circ}$ to around $60^{\circ}$, coherent dynamics is dominant, in which the well-known case of level repulsion [Fig.~\ref{Fig3}(a)] is obtained, with a splitting gap of $2 |g_{\rm eff}|$. It is worth noticing that, with tuning $\varphi$, the decreased field torque at $\omega_c$ leads to the decrease of $\tilde{g}_s$, while the dissipation $\tilde{g}_b$ due to the travelling waves gradually increases. At $\varphi$ around $60^{\circ}$, the radiative linewidth exchange point in Fig.~\ref{Fig3}(d) suggests that coherent coupling $\tilde{g}_s$ vanishes, causing the merging of energy levels [Fig.~\ref{Fig3}(b)]. Furthermore, when $\varphi$ is tuned greater than $\sim 60^{\circ}$, the dissipation due to the travelling waves overcomes the coherent coupling, with the inversion of the magnon linewidth contributed by standing and travelling waves can be observed. The advantage of the dissipation over the coherence can be clearly evidenced by level attraction in Fig.~\ref{Fig3}(c) with $\varphi=90^{\circ}$.

(II) The net result of coherence-dissipation competition ($\tilde{g}_s$-$\tilde{g}_b$ competition) as well as the transition from repulsion to attraction can be condensed into a single parameter---the effective coupling strength $g_{\rm eff}$. This parameter can be obtained by fitting the dispersion using Eq.~(3) in Ref.~\cite{MH18} [also Eq.~(\ref{transmission}) derived in the next section]. We note that, if no travelling wave is considered, the expected coupling strength $g_p$ should approach zero following a trigonometric function of $\varphi$\cite{supp} (see derivation in Supplementary Section~3). However, in fact, the existence of travelling wave counteracts the coherent process in a coupled system and results in a ``faster" drop to zero coupling at $\varphi$=$60^{\circ}$. Such zero coupling is situated in the radiative linewidth exchange point in Fig.~\ref{Fig3}(d), corresponding to the turning point from repulsion to attraction. Beyond $60^{\circ}$, the growing dissipation $\tilde{g}_b$ makes a major contribution to $g_{\rm eff}$. Hence we observed a uprising trend for the absolute coupling $|g_{\rm eff}|$. In the following section, we develop a quantitative theory to (i) quantitatively describe the competition between the coherence and dissipation via converting radiative linewidth to $\tilde{g}_s$ and $\tilde{g}_b$, and (ii) express the net result of competition $g_{\rm eff}$ in an analytical form, with the key result shown by Eq.~(\ref{mainfit}). With this theory, we can well reproduce the summarized  $g_{\rm eff}$ for each $\varphi$, as shown in Fig.~\ref{Fig3}(e).

\section{Theory: microscopic origin of level attraction}
We now set up a theoretical model to quantitatively describe our measurements. As revealed in observation, photon modes around cavity resonance are in a superposition of standing and travelling waves as demonstrated by magnon radiative linewidth, causing the failure of standard argumentation that is based on the coupling between magnons and pure standing waves. Nevertheless, integrating the travelling waves (continuous background) results in an effective coupling between a magnon and standing-wave component that allows us to use this standard argumentation.

\begin{figure} [!htbp]
	\begin{center}
		\epsfig{file=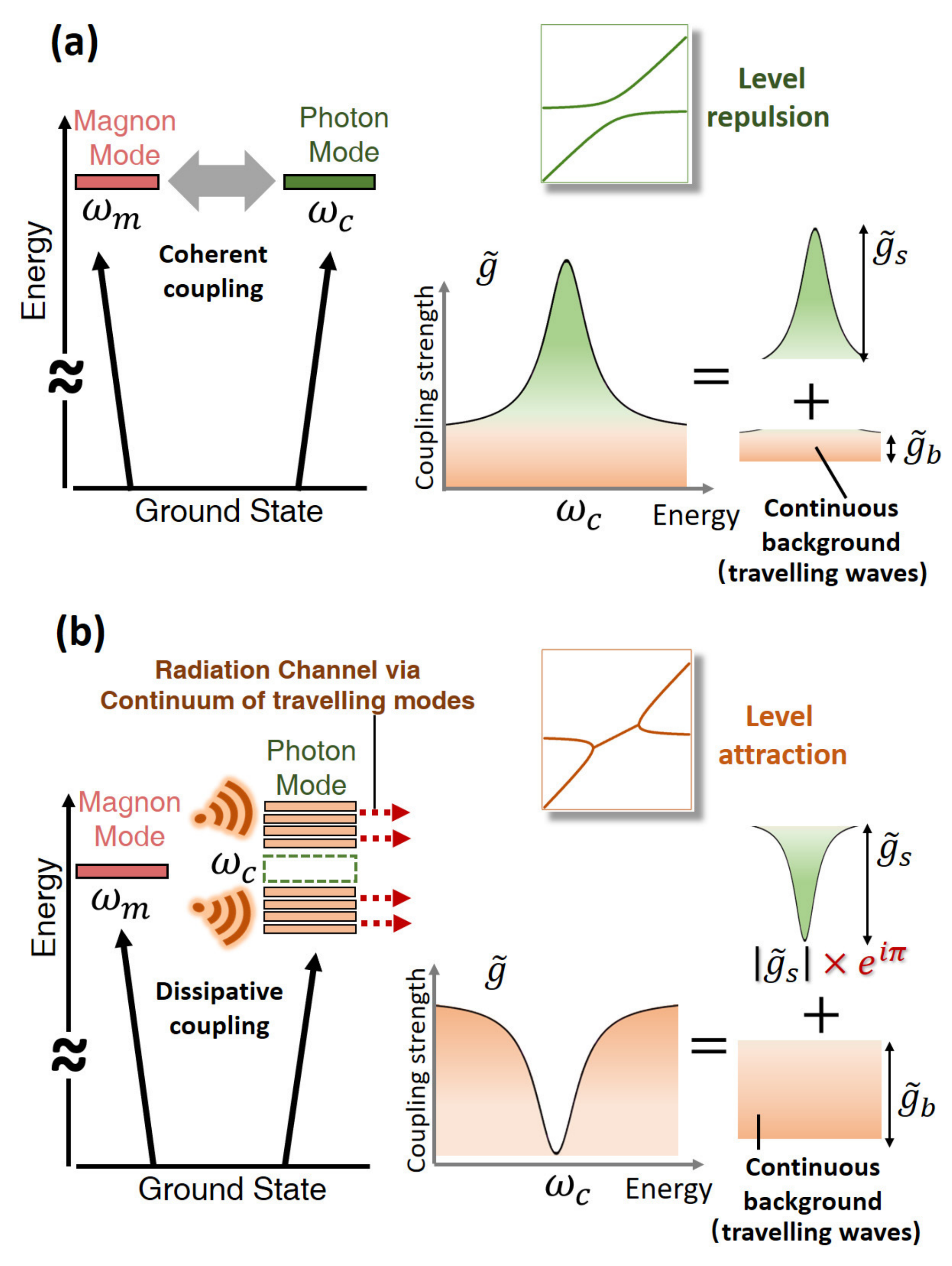,width=8.8cm}
		\caption{Coupling scheme and coupling strength profile. (a) Schematic figure of coherent magnon-photon coupling scheme and its coupling strength profile, which generates normal level repulsion effect. (b) As the continuum of travelling modes becomes dominant in the magnon-photon coupling, the level attraction can emerge.}
		\label{two_component}
	\end{center}
\end{figure}

With the photon modes in the 1D Fabry-Perot-like cavity parametrized by the momentum $k$ along the transmission direction, we begin the analysis of our system from the Fano-Anderson Hamiltonian \cite{Fano,Mahan}
\begin{equation}
\hat{H}=\omega_{\rm K}\hat{m}^{\dagger}\hat{m}+\sum_{k}\left(g_{k}\hat{m}^{\dagger}\hat{p}_{k}+
g^*_{k}\hat{m}\hat{p}^{\dagger}_{k}\right)\ ,
\label{original}
\end{equation}
in which $\hat{m}$ and $\hat{p}_{k}$ are magnon and photon operators, and $g_k$ denotes the coupling constant.
 The nature of the magnon-photon coupling can be addressed from the magnon Green's function, given by \cite{Fano,Mahan}
\begin{equation}
G_m(\omega)=\frac{1}{\omega-\omega_{\rm K}+i\delta_m-\Sigma(\omega)}\ ,
\end{equation}
in which $\delta_m$ represents the magnon intrinsic linewidth from Gilbert damping, and the magnon self-energy reads 
$\Sigma(\omega)=
\int {d\omega_{k}}{|\tilde{g}(\omega_{k})|^2}/{(\omega-\omega_{k}+i0_+)}$,
where $\tilde{g}(\omega_{k})=g_{k}\sqrt{{\cal D}(\omega_{k})}$ with ${\cal D}(\omega_{k})$ being the global density of state of photon. 
When $\omega_{k}$ is away from the resonance frequency $\omega_c$, $\tilde{g}({\omega_{k}})$ tends to be only contributed by the pure travelling photon, i.e., $\tilde{g}_b({\omega_{k}})$, in the circular waveguide, which slowly changes with momentum; whereas across the resonance frequency, $\tilde{g}({\omega_{k}})$  shows a peak or a dip as measured by magnon linewidth spectra [refer to Figs.~2(b) and (c)]. 
Therefore, $\tilde{g}({\omega_{k}})$ is divided into the continuous background
(see flat part in Fig.~\ref{two_component}) and the standing-wave contribution by
$\tilde{g}(\omega_{k})=\tilde{g}_b(\omega_{k})+\big[\tilde{g}(\omega_{k})-\tilde{g}_b(\omega_{k})\big]
=\tilde{g}_b(\omega_{k})+\tilde{g}_s(\omega_{k})$,
where we single out the contribution from the standing wave $\tilde{g}_s(\omega_{k})\equiv \tilde{g}(\omega_{k})-\tilde{g}_b(\omega_{k})$, which can be positive or negative (assuming $\tilde{g}_b$ is positive). 

With $\tilde{g}_s$ and $\tilde{g}_b$, the behaviour of $\tilde{g}({\omega_{k}})$  is schematically summarized in Fig.~\ref{two_component}, which can be divided by the contributions of the travelling-wave and standing-wave photon. Particularly, when the peak in $\tilde{g}({\omega_{k}})$ evolves to a dip, there is a sign change in $\tilde{g}_s$, corresponding to a phase jump $\pi$ relative to the continuous background, in the standing-wave contribution.

By definition of $\tilde{g}_s$ and $\tilde{g}_b$, the magnon self energy becomes
$\Sigma(\omega)
=-i\Gamma(\omega)+I(\omega)$
with $I(\omega)\equiv\int {d\omega_{k}}\frac{|\tilde{g}_s(\omega_{k})|^2+2e^{i\Phi}|\tilde{g}_s(\omega_{k})\tilde{g}_b(\omega_{k})|}
{\omega-\omega_{k}+i0_+}$,
where $\Gamma(\omega)=i\int{d\omega_{k}}{|\tilde{g}_b(\omega_{k})|^2}/(
{\omega-\omega_{k}+i0_+})\approx\pi|\tilde{g}_b(\omega)|^2$ by neglecting the small energy shift of a magnon due to the continuous background, and $\Phi=0$ ($\pi$) when $|g({\omega_{k_z}})|$ shows a peak (dip) at resonance. In the self-energy, $\Sigma(\omega)+i\Gamma$ can be treated to be the effective self-energy between a magnon and a standing-wave photon when integrating the travelling-wave contribution.

Around the resonance energy, the standing-wave mode behaves as a single energy level with the central energy $\omega_c$ and the broadening $\gamma_s$. With this understanding,  when $|\omega-\omega_c|\ll \gamma_s$, we establish the following correspondence
% (see Supplementary Material for derivation)
$I(\omega)\approx \pi\gamma_s \frac{|\tilde{g}_s(\omega_c)|^2+2e^{i\Phi}|\tilde{g}_s(\omega_c)\tilde{g}_b(\omega_c)|}
{\omega-\omega_c+i\gamma_s}$.
Thus, after integrating the continuous background, the Green's function of magnon becomes
\begin{equation}
G_m(\omega)=\Big\{\omega-\omega_m+i(\delta_m+\Gamma)-\frac{g_{\rm eff}^2}{\omega-\omega_c+i\gamma_s}\Big\}^{-1} \ ,
\label{Green_magnon}
\end{equation}
where we define
\begin{align}
g^2_{\rm eff}=\pi\gamma_s\left(|\tilde{g}_s(\omega_c)|^2+2e^{i\Phi}|\tilde{g}_s(\omega_c)\tilde{g}_b(\omega_c)|\right) \ .
\label{mainfit}
\end{align}
It is noted that $g^2_{\rm eff}$ can be positive, zero and negative. Particularly, it vanishes when $\tilde{g}_s=0$, and can be positive (negative) when $\tilde{g}_s$ and $\tilde{g}_b$ have the same (opposite) sign, which explains the turning point, level repulsion and attraction in our measurements. At the resonance with $\omega=\omega_{\rm K}=\omega_c$, the linewidth broadening of magnon is calculated to be
$\gamma\mu_0\Delta H=\delta_m+\Gamma+g^2_{\rm eff}/\gamma_s=\delta_m+\pi(\tilde{g}_s+\tilde{g}_b)^2$. While detuned away from the cavity resonance frequency, $\gamma\mu_0\Delta H\approx \delta_m+\pi\tilde{g}_b^2$. Thus, from the resonant and detuned measurements on the radiative linewidth broadening, one can obtain information of $\tilde{g}_{s}$ and $\tilde{g}_{b}$, as we described in the previous sections. We note that when reproducing the experimental values $g^2_{\rm eff}$, $\gamma_s$ is obtained from extracting cavity linewidth, and $\tilde{g}_{b,s}$ are increased by 10$\%$ accounting for parasitic deviations.

%With the above understanding, when we focus on the coupling between a magnon and a standing-wave component of the photon, 
Furthermore, we can write the model effectively by the non-Hermitian Hamiltonian
\begin{align}
\nonumber
\hat{H}_{\rm eff}&=(\omega_{\rm K}-i\delta_m-i\Gamma)\hat{m}^{\dagger}\hat{m}+(\omega_c-i\gamma_s)\hat{p}^{\dagger}_s\hat{p}_s\\
&+(|g_{\rm eff}|e^{i\Phi}\hat{m}^{\dagger}\hat{p}_s+|g_{\rm eff}|\hat{m}\hat{p}^{\dagger}_s) \ ,
\label{effective}
\end{align}
in which $\hat{p}_s$ represents the standing-wave photon operator, and the dissipative magnon-photon coupling is obtained when $\Phi=\pi$. This Hamiltonian can exactly recover the magnon Green's function [Eq.~(\ref{Green_magnon})], and hence provides an effective description of our system. All the parameters in this Hamiltonian can be either measured or calculated\cite{supp} (see Supplementary Section~4).
We use the input-output theory \cite{S_input_output2} to calculate the photon transmission,
\begin{equation}
S_{21}(\omega)=1-\frac{\gamma_s}{-i(\omega-\omega_c)+\gamma_s+\frac{g^2_{\rm eff}}{-i(\omega-\omega_{\rm K})+\delta_m+\Gamma}}\ .
\label{transmission}
\end{equation}
When there is no magnon coupled to the cavity mode, $g_{\rm eff}$ vanishes with $|S_{21}|$ showing no transmission at $\omega_c$. The existence of the magnon thus changes the behaviour of transmission, showing either level repulsion or attraction depending on the sign of $g_{\rm eff}^2$.

\begin{figure} [!htbp]
	\begin{center}
		\epsfig{file=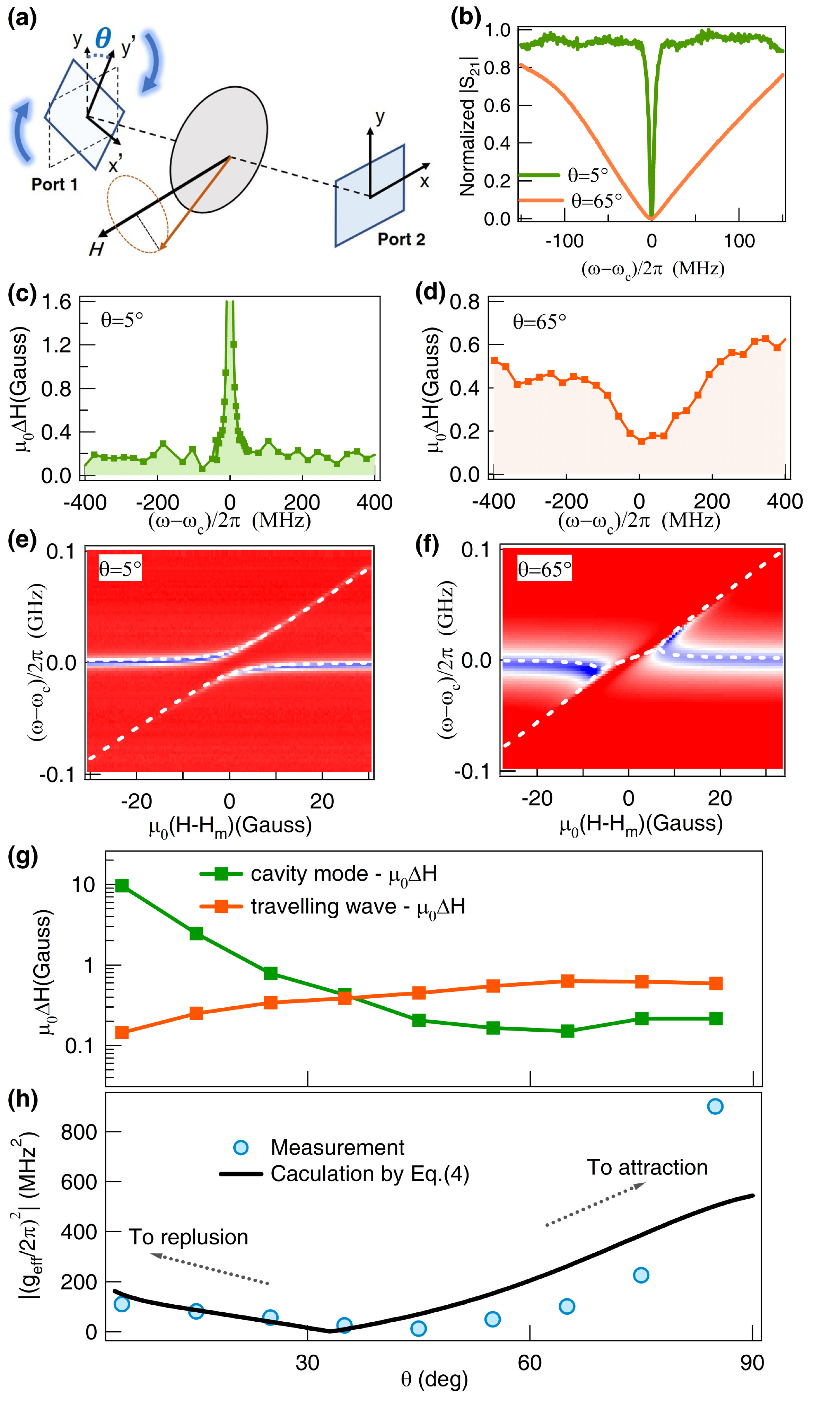,width=9cm}
		\caption{Observation of level attraction via tuning global electromagnetic environment. (a) Schematic figure of rotating the relative angle $\theta$ between two transitions. (b) The cavity linewidth broadening when tuning $\theta$ from $5^{\circ}$ to $65^{\circ}$. (c) and (d) Magnon radiative linewidth enhancement at cavity mode when setting $\theta$=$5^{\circ}$ and $65^{\circ}$. The level repulsion (e) and attraction (f) are obtained by setting $\theta$ to be $5^{\circ}$ and $65^{\circ}$, respectively, with red (blue) color denoting the maximal (minimal) signal. (g) Transition of the coherent and dissipative couplings as indicated by magnon radiative linewidth. The turning of effective coupling strength can be viewed in (h) when the coherent coupling evolves to the dissipative one.}
		\label{Fig5}
	\end{center}
\end{figure}

\section{Controlling level attraction via cavity geometry}
Theory proposed in the above section sets a general framework for describing the competition between coherent and dissipative dynamics. Based on this theory, more elements associated with dissipation can be included to realize level attraction, which is not only restricted to the manipulation of torque exerted by the field on magnetic materials. One natural way is to tune the weights of the travelling and standing waves, which is realizable by tuning the global geometry of our device. Here, we demonstrate an alternative method to transform level repulsion to attraction, namely by modification of the global electromagnetic environment. It may provide a playground for using photonic construction techniques, like metamaterials, to build level attraction. We can tune the global geometry of the waveguide cavity via rotating the relative angle $\theta$ between the two transitions, as we schematically show in Fig.~\ref{Fig5}(a). As a result, we enable to control the weight of the travelling wave (e.g., when $\theta=0^{\circ}$, our device behaves like a waveguide without standing waves), which is the key element for realization of the level attraction. This is evident by looking at the transmission spectrum for two different values of $\theta$, as shown in Fig.~\ref{Fig5}(b). By tuning $\theta$ from $5^{\circ}$ to $65^{\circ}$, the cavity mode linewidth $\gamma_s$ is significantly enhanced\cite{supp} (see $\theta$-dependence of $\gamma_s$ in Supplementary Section~5). This enhancement of the linewidth suggests that more travelling wave states are built in the cavity, through which more energy can be delivered to the environment. Thus, we construct the hybrid states (with $\varphi=70^{\circ}$) to realize the competition between the standing and travelling waves, and their coupling strength spectra can be well characterized by measuring the magnon radiative linewidth. Consistent with the information from the cavity mode linewidth, we found that the travelling wave component has turned the coupling strength profile from coherence dominant case [Fig.~\ref{Fig5}(c)] to dissipation dominant one [Fig.~\ref{Fig5}(d)]. Such modifications of the competition between coherence and dissipation further determine the dispersion of the magnon-photon hybrid system, as proved by the appearance of the transition from level repulsion to attraction [shown in Fig.~\ref{Fig5}(e) and (f), respectively], as we increase the dissipation by increasing $\theta$.

On a more detailed level, we plot the magnon radiative linewidth at the cavity resonance and travelling wave frequencies, respectively, to demonstrate the competition between the coherence and dissipation as we tune $\theta$ [see Fig.~\ref{Fig5}(g)]. When $\theta$ is tuned from $0^{\circ}$ to around $40^{\circ}$, Fig.~\ref{Fig5}(g) suggests that the coherence nature of the coupling is being counteracted by the growing dissipation through the travelling background. The net result in such competition is also reflected in the decreasing trend of the effective coupling strength $g_{\rm eff}$, as we show in Fig.~\ref{Fig5}(h). While for the angles greater than 40$^{\circ}$, Fig.~\ref{Fig5}(g) indicates that the growing dissipation exceeds the coherence, and has become the leading source in the effective coupling strength, resulting in $g_{\rm eff}$ to undergo an obvious increase as displayed in Fig.~\ref{Fig5}(h) when $\theta>40^{\circ}$. 

Quantitatively, $g_{\rm eff}$ can be reproduced by using Eq.~\eqref{mainfit} in our theoretical model, which further confirms that our theory catches the key feature of the coherence-dissipation engineering. In addition, we note that by tuning the global geometry here, the effective coupling strength in the level attraction can be obviously made larger than that in level repulsion, as can be seen by comparing $g_{\rm eff}$ when $\theta$=5$^\circ$ and 85$^\circ$. This is due to the increased broadening of cavity mode linewidth $\gamma_s$ when increasing $\theta$, which contributes to the enhancement of $g_{\rm eff}$ in magnitude [see in Eq.~\eqref{mainfit}]. Overall, in this part, by engineering the dissipation through tuning $\theta$, the obtained radiative linewidth characterization, the resultant dispersion transition from repulsion to attraction as well as the agreement of $g_{\rm eff}$ from both measurement and theory further confirm the physical understanding proposed in our work.

%The net product in such competition can be described by the effective coupling strength $g_{\rm eff}$, shown in Fig. \ref{Fig5}(h). When $\theta$ is tuned from $0^{\circ}$ to around $40^{\circ}$, the decrease of $g_{\rm eff}$ reflects the coherent nature of the coupling being counteracted by the growing dissipation, which is also presented in the form of {\em the} magnon radiative linewidth in Fig. \ref{Fig5}(g). 

\section{CONCLUSIONS}

Getting access to a full tunability of coupled states is a great challenge of modern physics, which requires control of both coherent and dissipative dynamics as well as of the competition between them. Benefiting from engineering such competition, the emerging phenomenon of level attraction dynamics has been considered on various physical platforms \cite{D3,MH18,BR,SSR,YY}. Our work brings three new perspectives for better understanding and utilizing the attraction dynamics, which are distinct from existing results with classical and phenomenological explanations. (i) Identifying the microscopic origin of dissipative coupling. Placing a YIG sphere in a waveguide cavity, we revealed travelling-wave state in the cavity as the microscopic origin of dissipative coupling in the level attraction. (ii) Providing a method to measure the coherent-dissipative coupling competition. With characterizing the magnon radiation, the magnon linewidth can quantify the coherent and dissipative coupling strengths. In this manner, we developed the theory to account for the effective coupling strength as a net result of competition.  (iii) Extending control dimensions in obtaining level attraction. By tuning the torque on magnetization or modifying the global photon environment, we enriched the toolbox of realizing level attractions. Our results open new ideas to utilize level attraction dynamics compared with existing techniques, and provide effective ways to explore level attraction in various harmonic systems.

%other control techniques with using optomechanical circuits \cite{BR}, adjusting resonator positions \cite{MH18} or applying two driving forces on a magnon \cite{XK}. Finally, we address that our revealed mechanism is not restricted to specific types of cavities such as Fabry-Perot cavity or split-rings.

\subsection*{ACKNOWLEDGMENTS}
This work was funded by the National Natural Science Foundation of China under Grant No.11804352, the Shanghai Pujiang Program No.18PJ1410600, the Science and Technology Commission of Shanghai Municipality (STCSM No. 16ZR1445400), SITP Innovation Foundation (CX-245) and Independent Research Project of State Key Laboratory of Infrared Physics (No. Z201920). T.Y. and Y.M.B. were supported by the Netherland Organization for Scientific Research (NWO). We would like to thank G.E.W.Bauer for useful discussions.

\subsection*{APPENDIX A: Measurement setup.}

The experimental apparatus in this work consists of a cylindrical waveguide as well as two circular-rectangular transitions. By further connecting two coaxial-rectangular adapters at both ends of our cavity, it enables the vector network analyzer (VNA) to pick up the transmission signal of our cavity. VNA sends microwave to the cavity, with an input power of 0~dBm. The function of circular-rectangular transition is that, it smoothly transforms the TE$_{10}$ mode of its rectangular port to the co-polarized TE$_{11}$ mode of its circular port and reciprocally. TE$_{10}$ waves entering the transitions from port 1 is transmitted through with almost zero reflection. When TE$_{11}$ microwaves exiting from the circular waveguide to the transitions, only the transformed TE$_{10}$ part can be transmitted without reflection, while the rest part of microwaves are almost totally reflected. As a result, such reflection of microwaves at both ends leads to the generation of standing wave in the long dimension of circular waveguide. 

\subsection*{APPENDIX B: Sample description.}
The YIG sphere is highly polished, with a diameter of 1~mm, mounted in the middle of our cavity via a scotch tape. Gilbert damping parameter of YIG sphere is around 4.5$\times$10$^{-5}$ with the measured inhomogeneous broadening at zero frequency being 0.19~Gauss and the saturated magnetization is $\mu_0 M_s=0.175~T$. By tuning external magnetic field, the resonance frequency of YIG sphere ($\omega_{\rm K}$) can be tuned, following the expression $\omega_{\rm K}=\gamma(H+H_A)$. Here, $\gamma$=180$\mu_0$~GHz/T is the electron gyromagnetic ratio, $\mu_0 H_A$=8.2~mT is the effective field and $\mu_0$ is the permeability of vacuum.

%\paragraph*{\textbf{Theory.}} In Supplementary Section~4, details about the theory derivation of the effective coupling strength and its numerical calculation are provided.  


\begin{widetext}

\section{Supplementary Section S1.\\ Characterization of magnon intrinsic damping.}

In this part, we characterized the intrinsic damping of magnon mode. The YIG sphere is placed in the center of a standard rectangular waveguide, and we connect the rectangular waveguide to the VNA (vector network analyzer) by a rectangular-to-coaxial adapters. By applying and tuning the microwave frequency ($\textbf{h}$) and the external static magnetic field ($H$) on YIG sphere, we can measure the response of magnon mode $|S_{21}(H)|$, with the typical results measured around 12~GHz shown in the Fig.~\ref{Fig1}(a) as an example. In this example, by applying lineshape fitting to the measured data, the magnon linewidth can be obtained with $\mu_0 \Delta H=0.36$~Gauss. Furthermore, by measuring the transmission $|S_{21}(H)|$ for each frequency as well as fitting their linewidth, we can obtain the magnon linewidth over a broad frequency range, as shown in the Fig.~\ref{Fig1}(b). By fitting the measured data using the relation $\mu_0 \Delta H=\alpha\omega/\gamma+\mu_0 \Delta H_0$ \cite{Fit}, we can obtain the Gilbert damping coefficient $\alpha$ as $4.5\times 10^{-5}$ and inhomogeneous broadening $\mu_0 \Delta H_0$ as 0.19~Gauss.

\begin{figure*} [!htbp]
	\begin{center}
		\epsfig{file=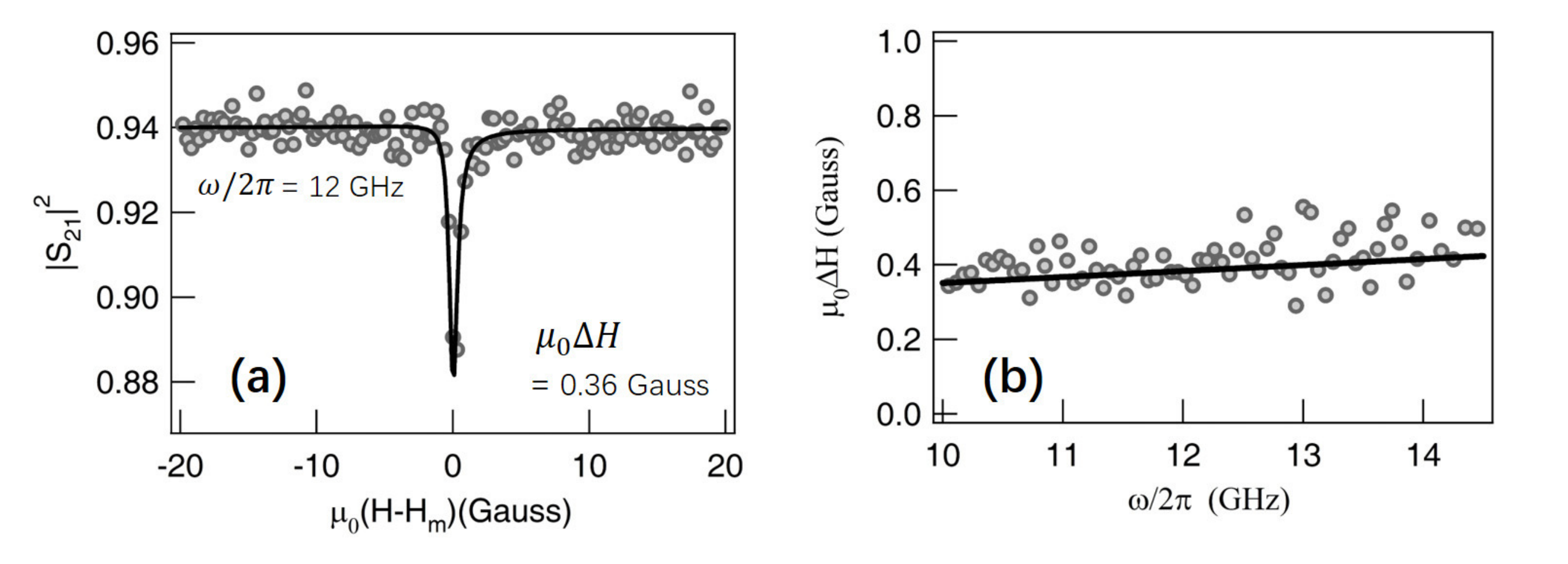,width=16.5cm}
		\caption{Measurement of the magnon linewidth with a standard rectangular waveguide. (a) Magnon resonance can be observed from measuring the photon transmission signal. Typically, we display here the magnon mode around $\omega/2\pi=12$~GHz, with the magnon linewidth $\mu_0 \Delta H$ can be obtained by lineshape fitting. (b) The Gilbert damping coefficient and inhomogeneous broadening at zero frequency can be obtained by applying linear fitting to $\mu_0 \Delta H_0-\omega$ relation. Circles and solid line are the measured data and fitted data, respectively.} 
		\label{Fig1}
	\end{center}
\end{figure*}

\section{Supplementary Section S2.\\ Magnon linewidths at cavity resonance and detuned frequency.}

Here, we display the measured signal $|S_{21}(H)|$ at both cavity resonance ($\omega_c$) and detuned frequencies ($\omega_d$). First, for case I with relative angle $\varphi$ between microwave magnetic field and external static field set to be $0^{\circ}$, we show the typical $|S_{21}(H)|^2$ at $\omega_c$ and at $\omega_d$ that is red-detuned 400~MHz away the resonance, as shown in Fig.~\ref{Fig2}(b) and (a), respectively. We obtain the magnon linewidth $\mu_0 \Delta H$ by fitting $|S_{21}(H)|^2$, as a result it can be found that magnon linewidth shows an enhancement at $\omega_c$. However, for case II with $\varphi$ set to be $90^{\circ}$, we observe a larger magnon linewidth at continuous wave range and a linewidth suppression at cavity resonance, as shown in Fig.~\ref{Fig2}(c) and (d). 

\begin{figure*} [!htbp]
	\begin{center}
		\epsfig{file=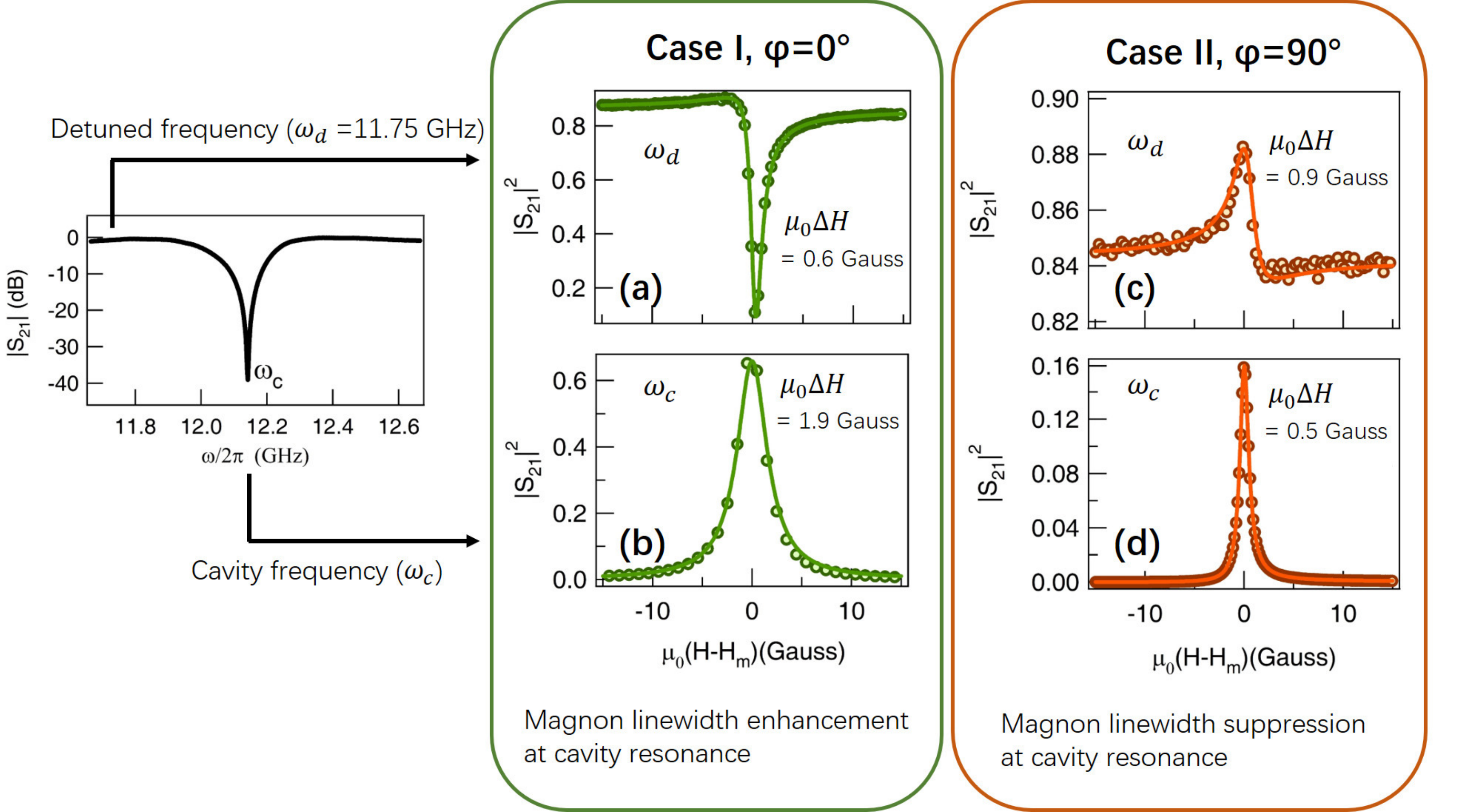,width=15cm}
		\caption{Broadband Magnon linewidths. Magnon linewidths at cavity resonance and detuned frequencies from the measured  $|S_{21}|^2$ as a function of the static magnetic field $H$. For Case I (II) with $\varphi=0^{\circ}$ ($\varphi=90^{\circ}$), we compare the magnon linewidth at detuned resonance and cavity frequencies in (a) and (b) [(c) and (d)].} 
		\label{Fig2}
	\end{center}
\end{figure*}

\section{Supplementary Section S3.\\ Dependence of coupling strength on microwave polarization.}

Here, we provide the derivation on the magnetic-field direction depenence of the magnon-photon coupling when neglecting the continuous waves.
In a well-confined cavity with no continuous waves considered, by the Zeeman energy the coupling strength $g_p\propto {\bf h}({\bf r})\cdot{\bf H}$,  where ${\bf h}({\bf r})$ is the local field intensity at the position of YIG sample \cite{HT}.
As we tune the relative angle $\varphi$ between ${\bf h}({\bf r})$ and $H$, $g_p$ varies. Assuming $\varphi=0^{\circ}$, the microwave magnetic field for the magnetic sphere to feel is the largest, referred to ${\bf h}_m({\bf r})$. In this situation, ${\bf H}$ is parallel to ${\bf h}_m({\bf r})$. Thus, when we tune the direction of ${\bf H}$ \cite{HT},  
\begin{align}
g_p\propto |{\bf h}_m({\bf r})|H\cos\varphi.
\end{align}
When $\varphi=90^{\circ}$, the magnon-photon coupling strength should be zero if there is no continuous waves.

\section{Supplementary Section S4.\\Derivation on the effective coupling strength}

Here we provide the detailed derivation for the effective coupling strength in $I(\omega)$ defined in the theoretical model. In the derivation of the self-energy, we approximately describe the distribution of $|\tilde{g}_s|$ by a Lorentzian
\begin{equation}
|\tilde{g}_s(\omega_{k})|=|\tilde{g}_s(\omega_c)
|\frac{\gamma_s^2}{(\omega_{k}-\omega_c)^2+\gamma^2_s}\ ,
\end{equation}
with which we calculate
\begin{eqnarray}
\nonumber
I(\omega)&&\approx  \int_{-\infty}^{\infty}{d\omega_{k}}\frac{|\tilde{g}_s(\omega_c)|^2}{(\omega-\omega_{k})+i0_+}
\Big[\frac{\gamma_s^2}{(\omega_{k}-\omega_c)^2+\gamma^2_s}\Big]^2+\int_{-\infty}^{\infty}{d\omega_{k}}
\frac{2e^{i\Phi}|\tilde{g}_s(\omega_c)\tilde{g}_b(\omega_c)|}{(\omega-\omega_{k})+i0_+}
\frac{\gamma_s^2}{(\omega_{k}-\omega_c)^2+\gamma^2_s}\\
&&=\pi\frac{\gamma_s}{2}\frac{|\tilde{g}_s(\omega_c)|^2}{\omega-\omega_c+i\gamma_s}\Big[\frac{i\gamma_s}{(\omega-\omega_c)+i\gamma_s}+1\Big]+\pi\gamma_s
\frac{2e^{i\Phi}|\tilde{g}_s(\omega_c)\tilde{g}_b(\omega_c)|}{(\omega-\omega_c)+i\gamma_s}\ ,
\end{eqnarray}
where we have extended the integral to infinity due to the decay behavior of Lorenzian when away from $\omega_c$.
When  $|\omega-\omega_c|\ll \gamma_s$, we arrive at 
\begin{equation}
I(\omega)\approx \pi{\gamma_s}\frac{|\tilde{g}_s(\omega_c)|^2+2e^{i\Phi}|\tilde{g}_s(\omega_c)\tilde{g}_b(\omega_c)|}{\omega-\omega_c+i\gamma_s}\ .
\end{equation}

To take into account the details relevant for the experiment, apart from direct experimental measurements, we can calculate $\tilde{g}({\omega_{k}})$ by using a Maxwell's equations solver for the waveguide with a complicated geometry. With the simulation, we also compared our theoretical calculation with the measurements, and obtained good agreements.

\section{Supplementary Section S5.\\ Characterization of cavity linewidth.}

\begin{figure*} [!htbp]
	\begin{center}
		\epsfig{file=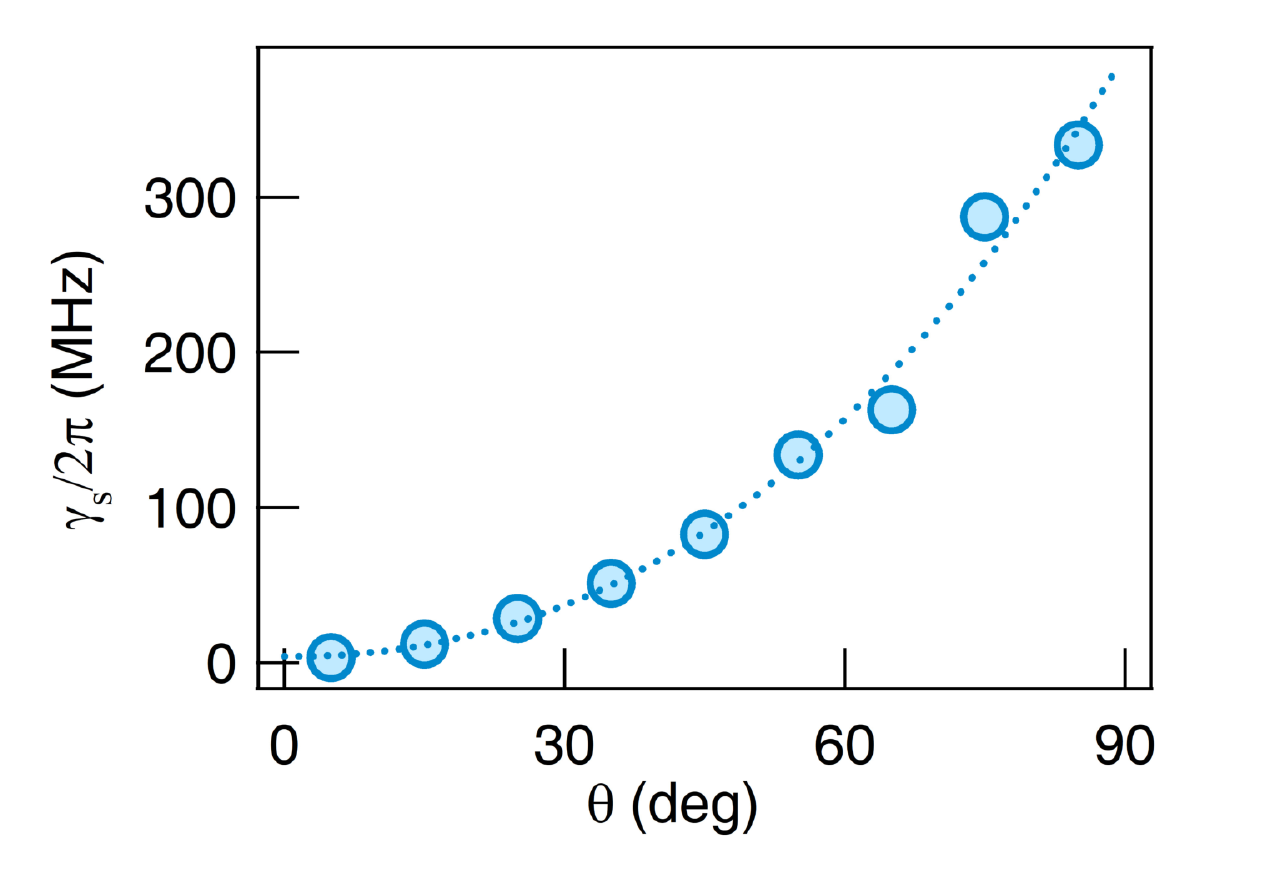,width=8.5cm}
		\caption{$\theta$-dependence of cavity linewidth. By tuning the relative angle $\theta$ between the two transitions in our device, we can observe the evolution of cavity linewidth $\gamma_s$ as shown by blue circles. Blue dashed line is a guidance to the eye.} 
		\label{Fig3}
	\end{center}
\end{figure*}

Here we characterized the cavity linewidth $\gamma_s$ by tuning the relative angle $\theta$ between the two transitions in our device. By inserting the metallic rotating parts in the center of our cavity, it enables us to continuously tune $\theta$, which also extends the length of our cavity leading to a slightly red-shift of cavity resonance frequency. By fitting the linewidth of transmission signal $|S_{21}(\theta)|$, we can obtain the cavity linewidth $\gamma_s$ as a function of $\theta$. We show in Fig.~\ref{Fig3} that, as tuning $\theta$ to a  larger value, the cavity linewidth is increased accordingly with more energy being able to dissipate to the outside environment.

\end{widetext}

\end{document}